\documentclass[superscriptaddress,showpacs,amsmath,amssymb,aps,prd,twocolumn,floatfix,nofootinbib]{revtex4-1}

\usepackage{graphicx}
\usepackage{dcolumn}
\usepackage{bm}
\usepackage{color}
\usepackage{amsmath}
\usepackage{amssymb}
\usepackage{lineno}
\usepackage{hyperref}
\usepackage{todonotes}
\usepackage{hhline}
\hypersetup{
    colorlinks = true
}

\usepackage{graphicx}

\begin{document}
\title{Scalable Deep Convolutional Neural Networks for Sparse, Locally Dense Liquid Argon Time Projection Chamber Data}


\newcommand{\SLAC}{SLAC National Accelerator Laboratory, Menlo Park, CA, 94025, USA}
\newcommand{\Stanford}{Stanford University, Stanford, CA, 94305, USA}
\affiliation{\Stanford}
\affiliation{\SLAC}

\author{Laura~Domin\'e} \affiliation{\Stanford}\affiliation{\SLAC}
\author{Kazuhiro~Terao} \affiliation{\SLAC}

\collaboration{on behalf of the DeepLearnPhysics Collaboration}\email{contact@deeplearnphysics.org} \noaffiliation

\begin{abstract}
Deep convolutional neural networks (CNNs) show strong promise for analyzing scientific data in many domains including particle imaging detectors such as a liquid argon time projection chamber (LArTPC). Yet the high sparsity of LArTPC data challenges traditional CNNs which were designed for dense data such as photographs. A naive application of CNNs on LArTPC data results in inefficient computations and a poor scalability to large LArTPC detectors such as the Short Baseline Neutrino Program and Deep Underground Neutrino Experiment. Recently Submanifold Sparse Convolutional Networks (SSCNs) have been proposed to address this class of challenges.  We report their performance on a 3D semantic segmentation task on simulated LArTPC samples. In comparison with standard CNNs, we observe that the computation memory and wall-time cost for inference are reduced by a factor of 364 and 33, respectively, without loss of accuracy. The same factors for 2D samples are found to be 93 and 3.1 respectively. Using SSCN and public 3D LArTPC samples, we present the first machine learning-based approach to the reconstruction of Michel electrons, a standard candle for energy calibration in LArTPC due to their very well understood energy spectrum. We find a Michel electrons identification efficiency of 93.9\% and a 96.7\% purity. Reconstructed Michel electron clusters yield 95.4\% in average pixel clustering efficiency and 95.5\% in purity.
The results are compelling in showing the strong promise of scalable data reconstruction technique using deep neural networks for large scale LArTPC detectors.
\end{abstract}


\maketitle

\section{\label{sec:intro}Introduction}
Deep convolutional neural networks (CNNs) have become the standard machine learning (ML) technique in the fields of computer vision, natural language processing, and other scientific research domains~\cite{NatureDL}. Applications of CNNs are actively developed for neutrino oscillation experiments~\cite{UBPaper1,UBPaper2,UBNature}, including those that employ liquid argon time projection chambers (LArTPC). LArTPCs are a type of particle imaging detector which can make 2D or 3D images of charged particles' trajectories with a breathtaking resolution ($\sim$mm/pixel) over many meters of detection volume. Current and future neutrino oscillation experiments using LArTPCs include MicroBooNE~\cite{MicroBooNE}, Short Baseline Near Detector (SBND)~\cite{SBN}, ICARUS~\cite{ICARUS} and the Deep Underground Neutrino Experiment (DUNE)~\cite{DUNE}.  The active volumes of these experiments are respectively about 90 tons, 112 tons, 600 tons and 40,000 tons of liquid argon.


Particle trajectories in LArTPC data, many of which have the shape of 1D lines, are recorded in 2D or 3D matrix format with an approximate pixel resolution of 3~mm to 5~mm. Each image has millions to billions of pixels for large LArTPC detectors (e.g. MicroBooNE produces 80~mega-pixels images). Those trajectories are produced by ionization electrons, and are thin (a few pixels in width) and continuous. In each recorded data, depending on the experimental environment, there may be a few to dozens of particle trajectories. Therefore LArTPC images are generally sparse, yet locally dense (i.e. no gap in between pixels that form a trajectory). This characteristic of LArTPC data poses two serious challenges for the application of CNNs. First, the matrix algebra associated with CNNs is computationally inefficient for LArTPC data which is mostly filled with zeros. 
Second, in photographs for which CNNs are originally developed, all pixels carry information. The strength of CNNs to automatically extract signal features may be affected when applied on mostly zero-filled LArTPC data.



Recently a Submanifold Sparse Convolutional Network (SSCN)~\cite{SSCN1,SSCN2} has been proposed to address these concerns with data represented by sparse matrix or point clouds. In this paper we demonstrate that SSCN holds strong promise for analyzing LArTPC image data with respect to both accuracy and computational efficiency, thus for being scalable to future large detectors including DUNE. Our contributions include:
\begin{itemize}
    \item Demonstration of the better performance and scalability of sparse techniques in data reconstruction tasks that may be part of a general reconstruction chain in LArTPC data.
    \item Study of typical mistakes made by the algorithms and mitigation methods.
    \item First ML-based approach for the reconstruction of Michel electrons using a publicly available simulation sample, and quantification of the purity and efficiency of this approach.
\end{itemize}

All studies in this paper are reproducible using a {\sc Singularity} software container\footnote{\url{https://www.singularity-hub.org/containers/6596}}~\cite{Singularity}, our implementation of semantic segmentation algorithms\footnote{\url{https://github.com/Temigo/uresnet_pytorch}}, and public data samples\footnote{\url{https://dx.doi.org/10.17605/OSF.IO/VRUZP}}~\cite{PublicSample} provided and maintained by the DeepLearnPhysics collaboration. 

Section~\ref{sec:data} gives an overview of the public data set used in this paper. Section~\ref{sec:network} details the design of a neural network, U-ResNet, chosen for studying the impact of SSCN. Section~\ref{sec:experiments} describes our experiment including the performance metrics and training setup. Section~\ref{sec:results} presents the results including the performance comparison between a SSCN (sparse) and standard (dense) implementations of U-ResNet.  We discuss causes of poor performance of U-ResNet and propose mitigation methods in Section~\ref{sec:discussion}. Lastly, in  Section~\ref{sec:michel}, we present our approach and the results of reconstructing Michel electrons in the public simulation sample using the sparse U-ResNet.


\begin{figure*}[t]
    \centering
    \includegraphics[width=0.46\textwidth]{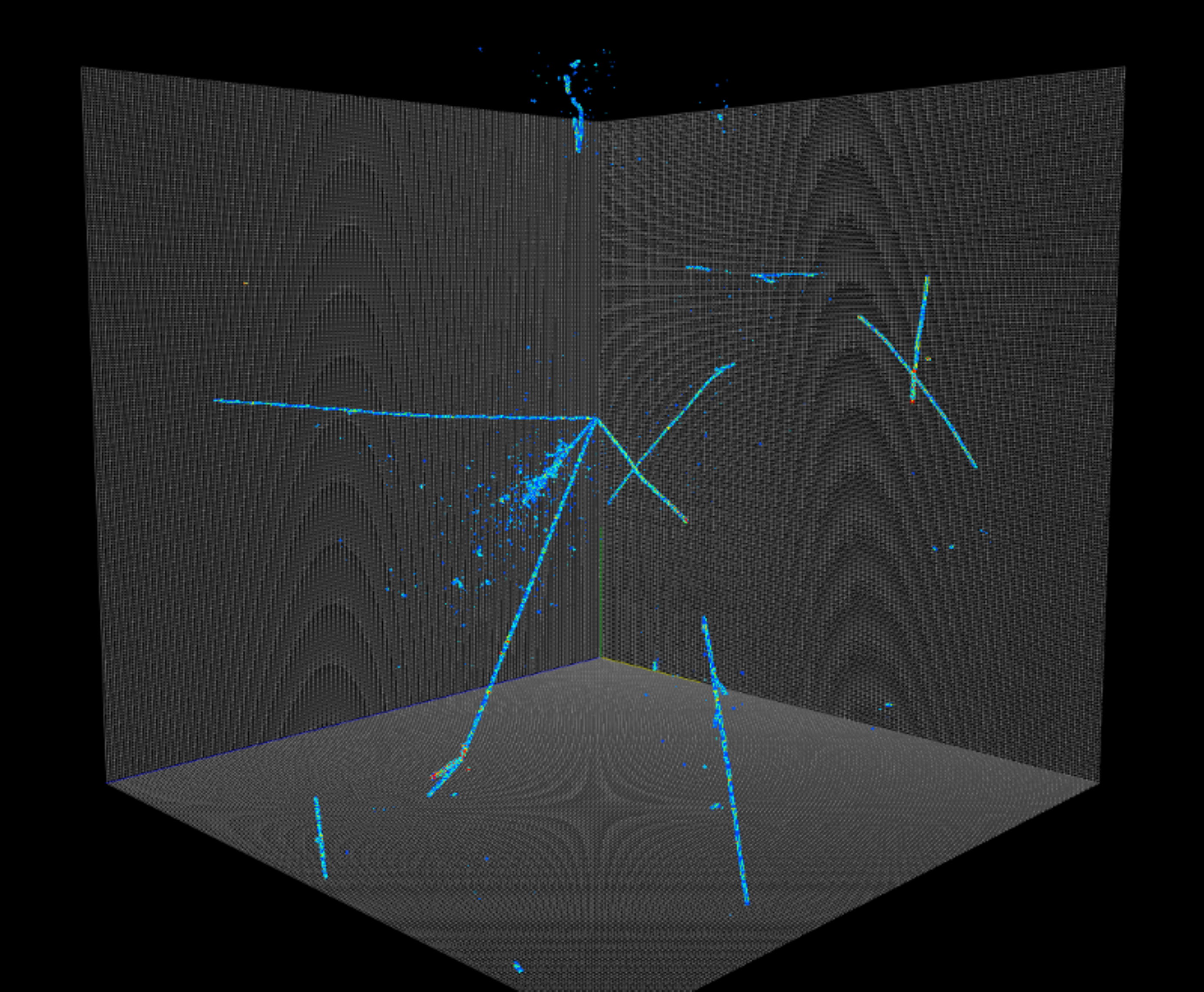}
    \includegraphics[width=0.45\textwidth]{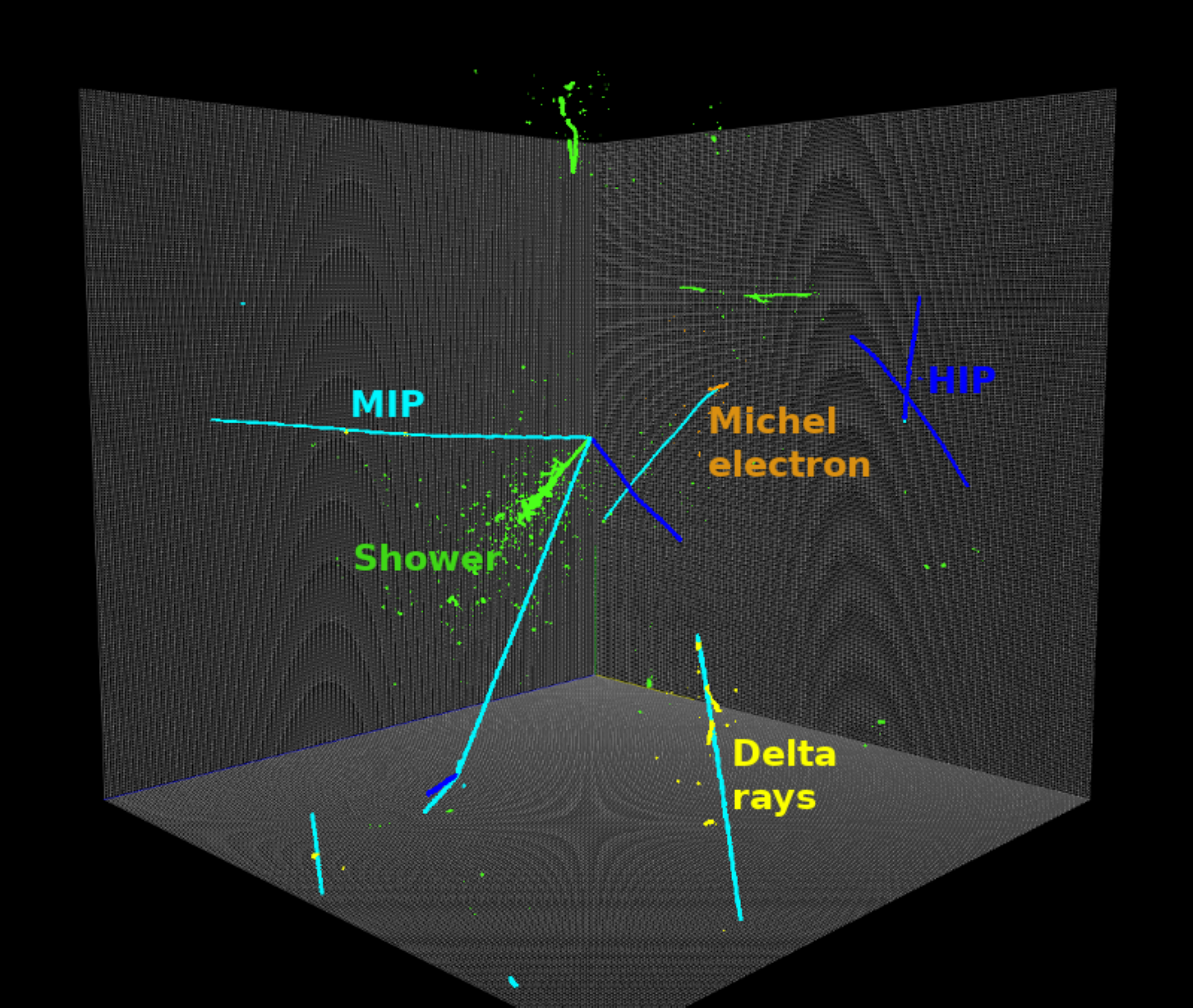}
        \caption{Simulated LArTPC event data (left) and labels (right). The data shows energy deposits from charged particle trajectories. The color corresponds to an energy scale. In the label image, each pixel is assigned one of five colors: heavily ionizing particles (HIP) in blue, minimum ionizing particles (MIP) in cyan, electromagnetic showers in green, delta rays in yellow and Michel electrons in orange. }
    \label{fig:lartpc}
\end{figure*}

\section{Data Set}
\label{sec:data}

\subsection{Particle Images}
In this paper we use 2D and 3D LArTPC simulation samples made publicly available by the DeepLearnPhysics collaboration~\cite{PublicSample}. These are images of particles traversing  a cubic volume of liquid argon, whose size can be $192$px, $512$px or $768$px. The spatial resolution of each pixel is 3~mm. The dataset contains 100,000 images for each size. We split each sample into 80~\% and 20~\% fractions as train and test sets respectively. There are two sources of particles in this dataset:
\begin{itemize}
    \item Single, isolated particle: an electron, muon, anti-muon or proton. 
    1 to 10 such particles are generated in a larger volume and a cropped 3D volume is recorded.
    \item Multi-particle vertex: multiples particles produced at the same 3D point, including electrons, gamma-rays, muons, anti-muon, charged pions, and protons.
\end{itemize}

Particle interactions with the liquid argon medium are simulated using Geant4~\cite{Geant4} and LArSoft~\cite{LArSoft}. The particle energy depositions are recorded in each pixel. The drift simulation, which would include for example the electron lifetime, recombination, diffusion and space charge effects, is not included in this dataset. However, energy depositions are smeared by a few pixels to mimic a diffusion effect while total deposited energy is conserved.

\subsection{Labels}
Among the tasks available for a benchmark in this public dataset, we choose the semantic segmentation.
The task is to predict a class of particle at pixel-level. The labels in the dataset for supervised learning include five possible classes for each pixel: 

\begin{table}[t]
    \centering
    \begin{tabular}{|l|c|c|c|c|c|}
        \hline
         & HIP & MIP & Shower & Delta rays & Michel $e^{-}$ \\
         \hline
        Fraction & 17~\% & 34~\% & 47~\% & 1~\% & 1~\% \\
        \hline
    \end{tabular}
    \caption{On average only 0.01~\% of pixels in an event are non-zero. This table shows to which class these non-zero pixels belong.}
    \label{tab:fractions}
\end{table}

\begin{itemize}
    \item Protons, referred to as heavily ionizing particle (\textbf{HIP}), which typically display short, highly ionized tracks
    \item Minimum ionizing particles (\textbf{MIP}) such as muons or pions, with usually longer tracks
    \item Electromagnetic \textbf{showers} induced by electrons, positrons and photons, with kinetic energy above critical value (about $33$MeV in argon)
    \item \textbf{Delta ray} electrons from hard scattering of other charged particles
    \item \textbf{Michel electrons} from the decay of muons
\end{itemize}

The statistics of each class in the dataset are shown in Table~\ref{tab:fractions}. Figure~\ref{fig:lartpc} shows an example of a simulated image from this dataset and the corresponding pixel-wise labels.
More details about the dataset can be found in the reference~\cite{PublicSample}.

\begin{figure*}[t]
    \centering
    \includegraphics[width=0.98\textwidth]{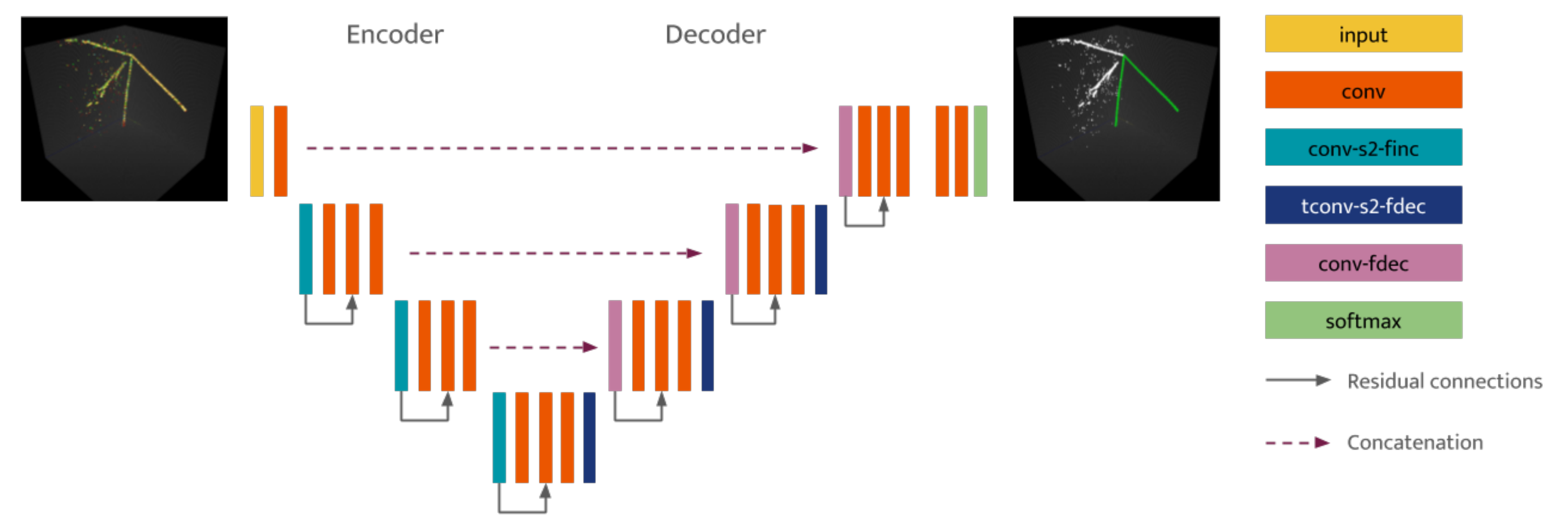}
    \caption{U-ResNet architecture for semantic segmentation. In this example we say that the U-ResNet has a depth of 3 since we perform 3 downsamplings. Turquoise boxes represent convolutions with stride 2 and increasing the number of filters. Dark blue boxes are transpose convolutions with stride 2 and decreasing the number of filters. Purple boxes are convolutions with stride 1 that decrease the number of filters. The spatial size of feature maps is constant across the horizontal dimension.}
    \label{fig:U-ResNet}
\end{figure*}

\section{Network architectures}
\label{sec:network}
\subsection{Dense U-ResNet: baseline}
We use a network architecture which we call U-ResNet. It is a hybrid between two popular architectures: U-Net~\cite{UNet} and ResNet~\cite{ResNet}.

U-Net is an auto-encoder network architecture (Figure~\ref{fig:U-ResNet}) which has been successful for medical image segmentation. It is made of two parts: the first half down-samples the spatial size (using strided convolutions in our case) the input image with several convolution blocks. This part learns the image features on different scales in a hierarchical manner, yielding a tensor with a low spatial resolution but a large number of channels. These channels contain a lot of compressed feature information: hence it is called the encoder part of the U-Net. The number of downsizing operations is referred to as {\it depth} in this paper, and affects the receptive field area of the network. The second half of the network applies to this tensor several up-sampling (we use transpose convolutions) and convolution operations. It is called a decoding path. Concatenation between the feature map of the previous layer in the decoding path and the feature map of the same spatial size in the encoding path helps the decoder to restore the original image resolution. The input image is single channel but the output has as many channels as there are classes.

U-Net is a generic CNN architecture. In our case each block of convolutions/up- or down- sampling is made of two convolution layers, followed by a batch normalization and a rectified linear unit (ReLU) function. According to the ResNet architecture we also add residual skip connections which allow the network to learn faster and be deeper. In our implementation the number of filters at each layer increases with depth in a power law for our dense U-ResNet and linearly for our sparse U-ResNet.

The strong performance of this network for 2-class semantic segmentation (between particle track and electromagnetic shower) at pixel-level on real detector data was already demonstrated~\cite{UBPaper2} by MicroBooNE experiment, which makes it a network of choice to benchmark a sparse technique on LArTPC simulation data.

\subsection{Submanifold Sparse Convolutional Networks}
The key element of submanifold sparse convolutional networks~\cite{SSCN1} is a so-called submanifold sparse convolution operation. It was designed for cases where the effective dimension of the data is lower than the ambient space, for example a 2D surface or a 1D curve in a 3D space. For such cases the standard dense convolutions are not suitable for several reasons:
\begin{itemize}
    \item Traditional convolutions involve dense matrix multiplication operations, which are computationally inefficient for sparse data.
    \item The submanifold dilation problem, as described in reference~\cite{SSCN1}: a single non-zero site in the image yields $3^d$ non-zero sites in the next feature map after a dense convolution, where $d$ is the spatial dimension (in our case $d=2$ or $d=3$). After 2 convolutions there will be $5^d$ non-zero sites, and so on. This inescapable growth ``dilates'' the originally sparse image which becomes denser. 
\end{itemize}

The idea of SSCNs is to keep the same level of sparsity throughout the network computations, especially convolutions. It has been shown to require significantly less computations while outperforming the dense CNNs on two 3D semantic segmentation challenges in the field of computer vision~\cite{SSCN2}.

Reference~\cite{SSCN1} defines two new operations. First, sparse convolutions $SC(n, m, f, s)$ with $n$ input features, $m$ output features, $f$ filters and a stride $s$ are defined. They address the first issue mentioned above and work in the same way as standard convolutions except they assume that the input from non-active pixels, which are zero or close to zero, is zero. The output feature map will have a size $(l-f+s)/s$ where $l$ is the size of the input. Secondly they define a submanifold sparse convolution $SSC(n, m, f)$ with similar notations as a modified $SC(n, m, f, s=1)$: the input is padded with $(f-1)/2$ zeros on each side to ensure that the output image will have the exact same size. An output pixel will be nonzero if and only if the central pixel of the receptive field is nonzero in the input feature map. $SSC$ operation tackles the second issue by constraining the output sparsity. In order to build complete CNNs based on these two operations, the authors also define a set of other custom operations such as activation functions and batch normalization layers by restricting the corresponding standard operations to the set of nonzero pixels.

\section{Experiments}
\label{sec:experiments}

We perform two sets of experiments. First, we compare the performance between dense and sparse U-ResNet using several evaluation metrics for 2D and 3D samples. Second, we study the variation of the performance of sparse U-ResNet with key architecture hyper-parameters and different image sizes.

\subsection{Evaluation Metrics}
The network is trained by minimizing a loss which is a softmax cross-entropy loss averaged over all the pixels of an image. We define different metrics of interest:
\begin{itemize}
    \item Non-zero accuracy: fraction of non-zero pixels whose label is correctly predicted.
    \item Class-wise non-zero accuracy: for each event and for each class, fraction of non-zero pixels in that class that are correctly predicted.
    \item Resources usage during the training and testing time:
    \begin{itemize}
        \item GPU memory occupied
        \item Computation wall time
    \end{itemize}
\end{itemize}

\subsection{Implementation and Training Details}
All networks were implemented using the PyTorch \cite{AutoGrad} (version 1.0) deep learning framework. SSCN relies on the library {\sc SparseConvNet} \footnote{\url{https://github.com/facebookresearch/SparseConvNet}}. We use {\sc LArCV2} \footnote{\url{https://github.com/DeepLearnPhysics/larcv2}} to interface with the LArTPC data files.
To train the networks we used ADAM optimizer~\cite{Adam} with the default learning rate of $0.001$. We trained the networks for 30k iterations in 3D and 40k iterations in 2D.
We used NVIDIA V100 GPUs with 32GB memory. On 3D images of size $192$px approximately 10 and 212 hours were required for convergence of the sparse and dense networks respectively.

\section{Results}
\label{sec:results}
\textbf{Notation}: we write, for example, [2D, 512px, 5-16] to represent ``2D images of size 512px, and U-ResNet of depth 5 with 16 filters''.

\subsection{Sparse vs dense U-ResNet}

We start by comparing the performance of dense versus sparse U-ResNet using the non-zero accuracy metric as well as the computational resources usages at train and inference (or test) time.

\begin{table}[t]
    \centering
    \begin{tabular}{|l|c|c|c|c|c|}
        \hline
        & Dense & \multicolumn{4}{c|}{Sparse} \\
        \hline
        Batch size & 4 & 4 & \multicolumn{3}{c|}{64} \\
        \hline
        Image size & 192px & 192px & 192px & 512px & 768px \\
        \hline
        Nonzero accuracy mean & 92\% & 94\% & 98\% & 99\%  & 99\%  \\
        \hline 
        Nonzero accuracy std & 0.096 & 0.088 & 0.049 & 0.014 & 0.0037 \\
        \hline
        \multicolumn{6}{|c|}{\textbf{Nvidia V100 GPU}}\\
        \hline
        Memory (test) [GB] & 16 & 0.044 & 0.19 & 0.67 & 1.3 \\
        \hline
        Memory (train) [GB] & 26x4 & 0.21 & 1.3 & 5.1 & 9.3 \\
        \hline
        Wall-time (test) [s] & 3.3 & 0.10 & 0.66 & 2.4 & 4.4 \\
        \hline
        Wall-time (train) [s] & 25 & 0.21 & 1.2 & 5.0 & 8.8 \\
        \hline
        \multicolumn{6}{|c|}{\textbf{Intel Xeon Silver 4110 CPU}}\\
        \hline
        Memory (test) [GB] & - & 0.57 & 0.81 & 1.9 & 3.0\\
        \hline
        Memory (train) [GB] & - & 0.59 & 1.9 & 3.9 & 4.0\\
        \hline
        Duration (test) [s] & - & 0.25 & 1.7 & 8.0 & 16 \\
        \hline
        Duration (train) [s] & - & 1.1 & 6.1 & 24 & 47 \\
        \hline
    \end{tabular}
    \caption{Sparse and dense U-ResNet scalability with the 3D image spatial size. The dense U-ResNet could not fit 3D images of size 512px nor 768px on a single GPU. Both sparse and dense networks here have a depth 5 and number of filters 16.}
    \label{tab:dense_vs_sparse}
\end{table}

As shown in Table~\ref{tab:dense_vs_sparse}, for a fixed 3D image size of 192px and identical training parameters (notably batch size), the final non-zero accuracy mean value over the whole dataset for the sparse U-ResNet is slightly higher than the dense counterpart by 2\%. However using the exact same architecture doesn't do justice to the real feat of SSCN: the GPU memory usage and computation duration are drastically cut down using sparse convolutions, which allows us to train the sparse U-ResNet with a dramatically larger batch size and a larger 3D image size. Harnessing both of these advantages allows one to beat the baseline dense 3D U-ResNet by a large margin in non-zero accuracy.  

\begin{figure}[t]
    \centering
    \includegraphics[width=0.48\textwidth]{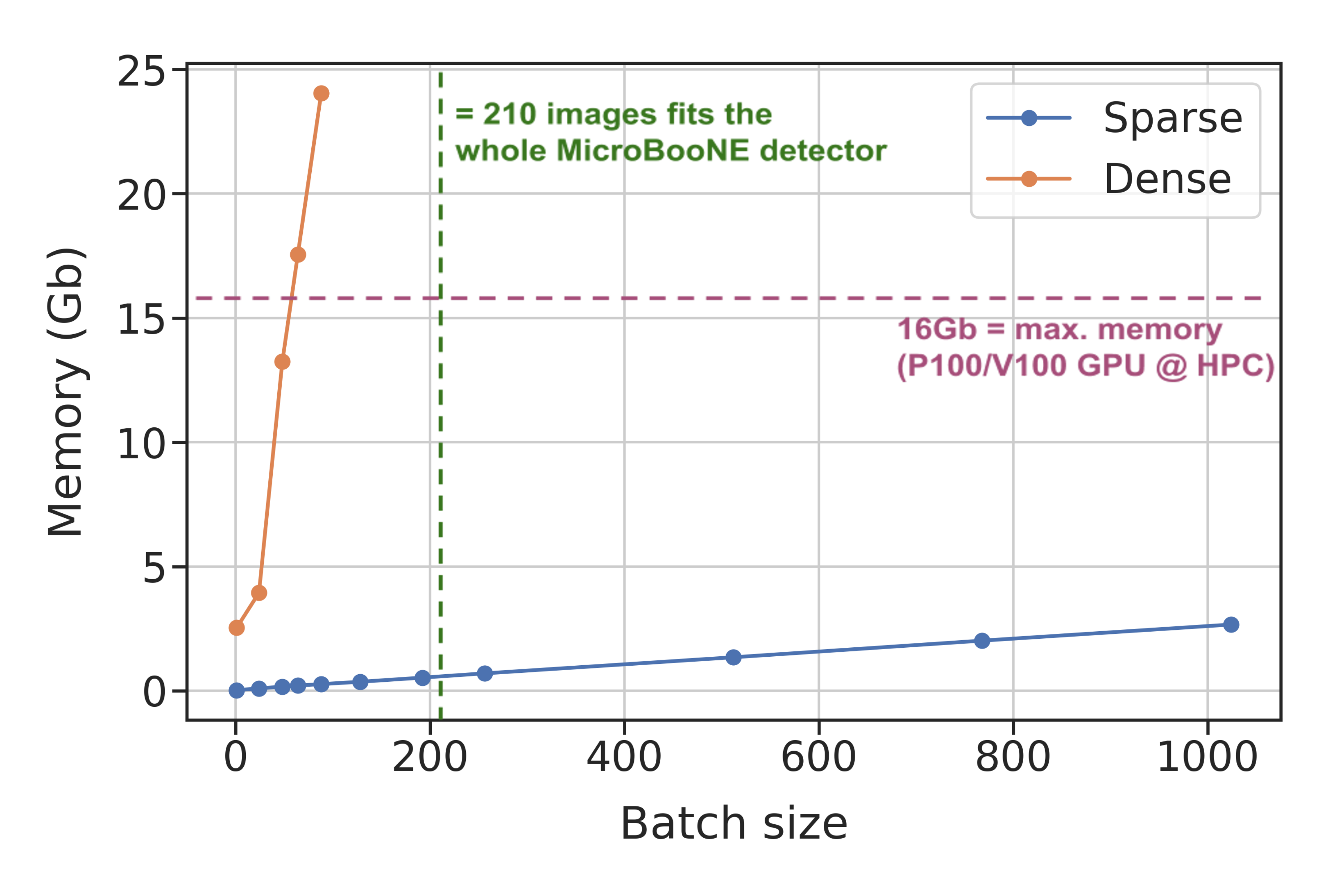}
    \caption{GPU memory usage as a function of batch size at inference time [2D, 512px, 5-16].}
    \label{fig:mem_bs}
\end{figure}

\begin{figure}[t]
    \includegraphics[width=0.48\textwidth]{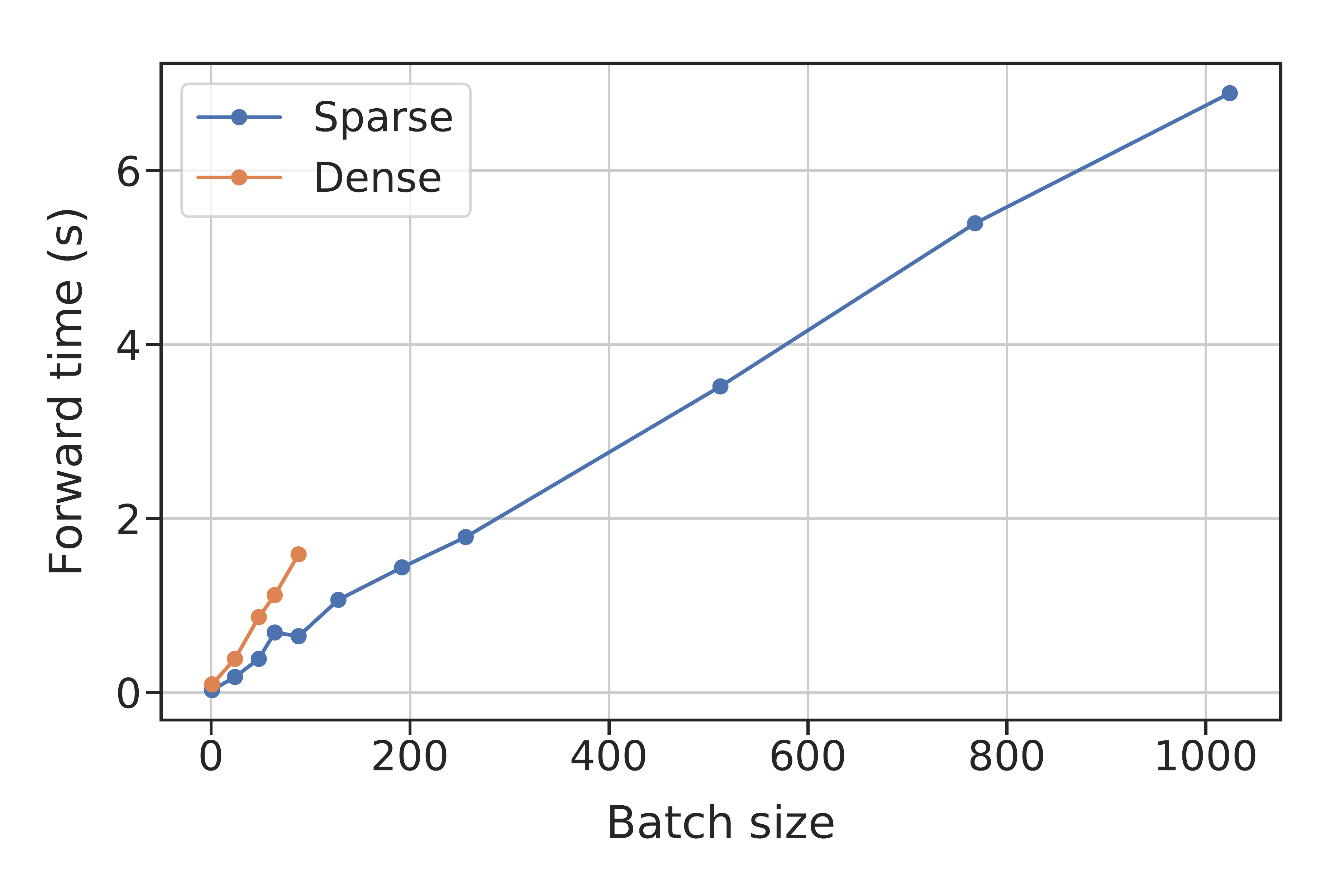}
    \caption{Computation wall-time as a function of batch size at inference time [2D, 512px, 5-16].}
    \label{fig:forward_bs}  
\end{figure}

\begin{figure}[t]
    \centering
    \includegraphics[width=0.48\textwidth]{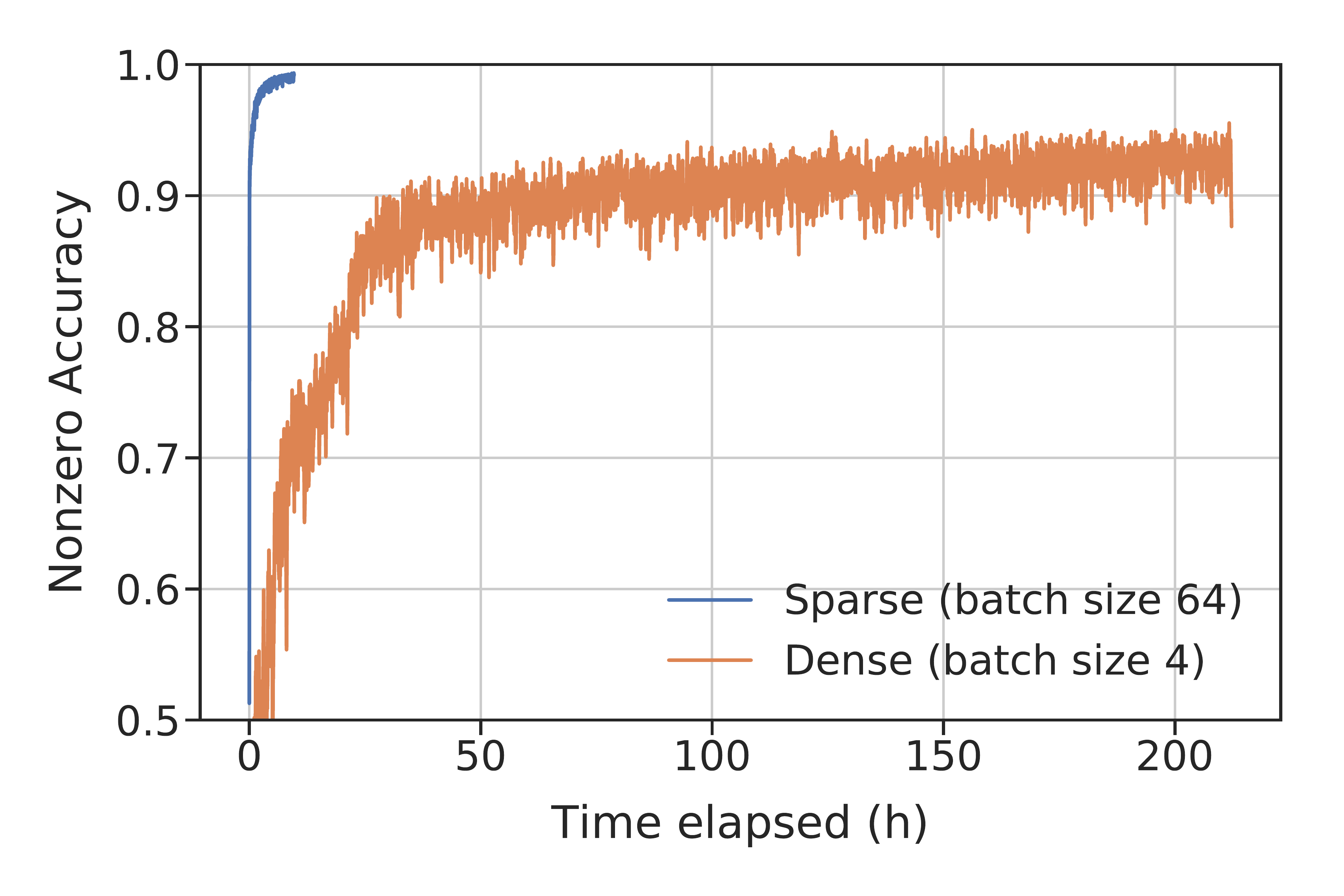}
    \caption{Nonzero accuracy as a function of wall-time during the training [3D, 192px, 5-16]. The sparse U-ResNet uses a batch size of 64, dense U-ResNet uses a batch size of 4.}
    \label{fig:acc_time}
\end{figure}

Figures~\ref{fig:mem_bs} and~\ref{fig:forward_bs} show the variation of memory and computation wall-time for sparse U-ResNet in [2D, 512px, 5-16].  The latter grows linearly but slowly as a function of batch size, which makes larger batch sizes practical not only for training but also for the inference. In particular, the sparse U-ResNet can easily process a whole MicroBooNE event data with a conventional GPU (memory of 4 to 11GB). The resource usage scales well with the batch size to handle ICARUS detector, which is about 6 times larger than MicroBooNE. At the batch size 88, which is the maximum possible for a single NVIDIA V-100 GPU with the dense version, the reduction factors for memory and computation wall-time with the sparse U-ResNet are 93 and 3.1 respectively. Further, because the computational cost scales with nonzero pixel count instead of the total pixel count in the bounded volume, sparse U-ResNet will be an ideal solution for DUNE far detector which will be sparser in the absence of cosmic rays. These benefits apply to a training phase of an algorithm.  Figure~\ref{fig:acc_time} shows how using sparse U-ResNet speeds up the training by several orders of magnitude. This is crucial for reconstruction algorithm R\&D work which often requires a short turn-around time for development.



\begin{figure}[t]
    \centering
    \includegraphics[width=0.48\textwidth]{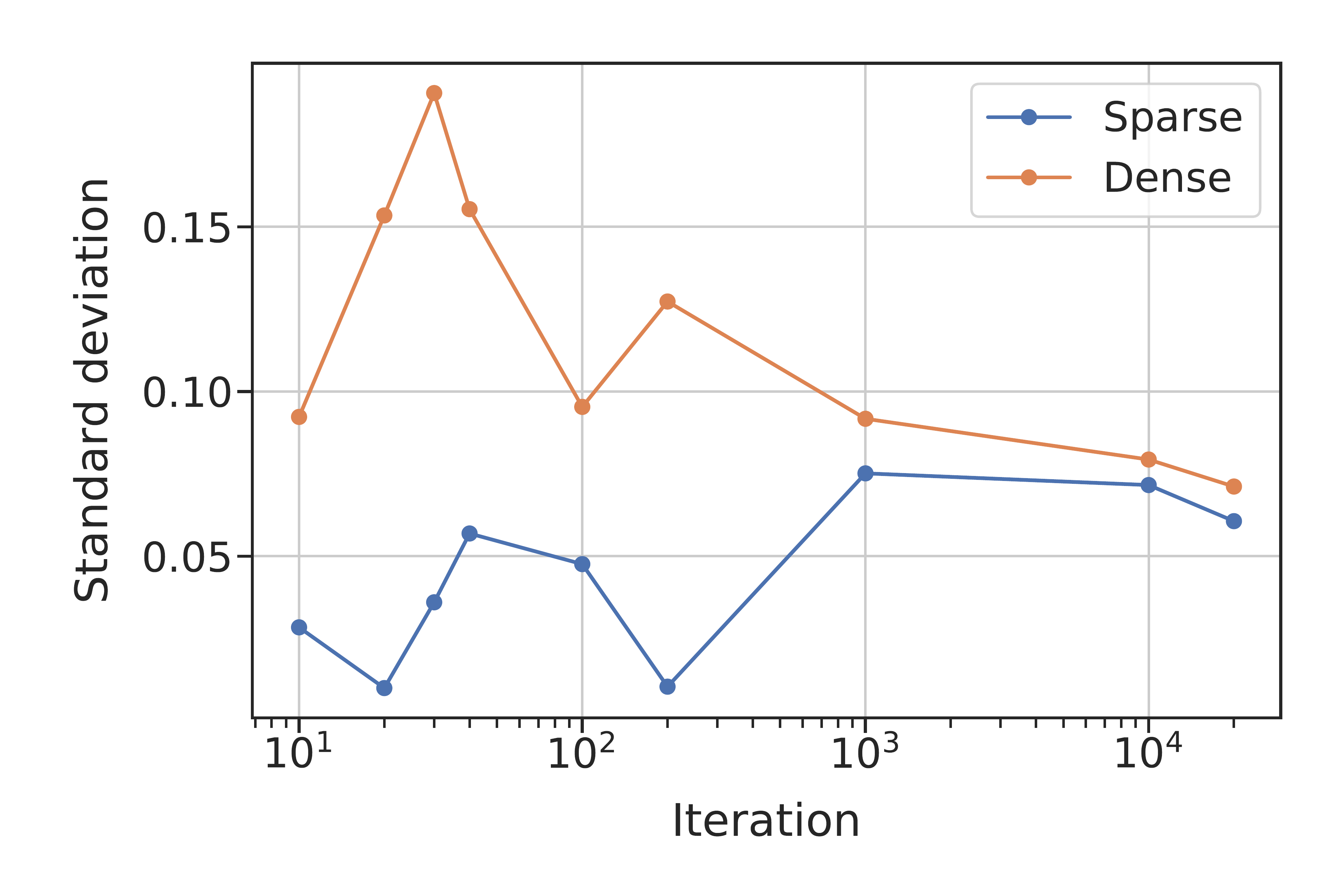}
    \caption{Standard deviation of the mean softmax value of pixels predicted as shower pixels in an image, as a function of the training iteration step. The sparse U-ResNet appears to learn in a more uniform manner across the pixels [2D, 512px, 5-16].}
    \label{fig:std}
\end{figure}

\begin{figure}[t]
    \centering
    \includegraphics[width=0.48\textwidth]{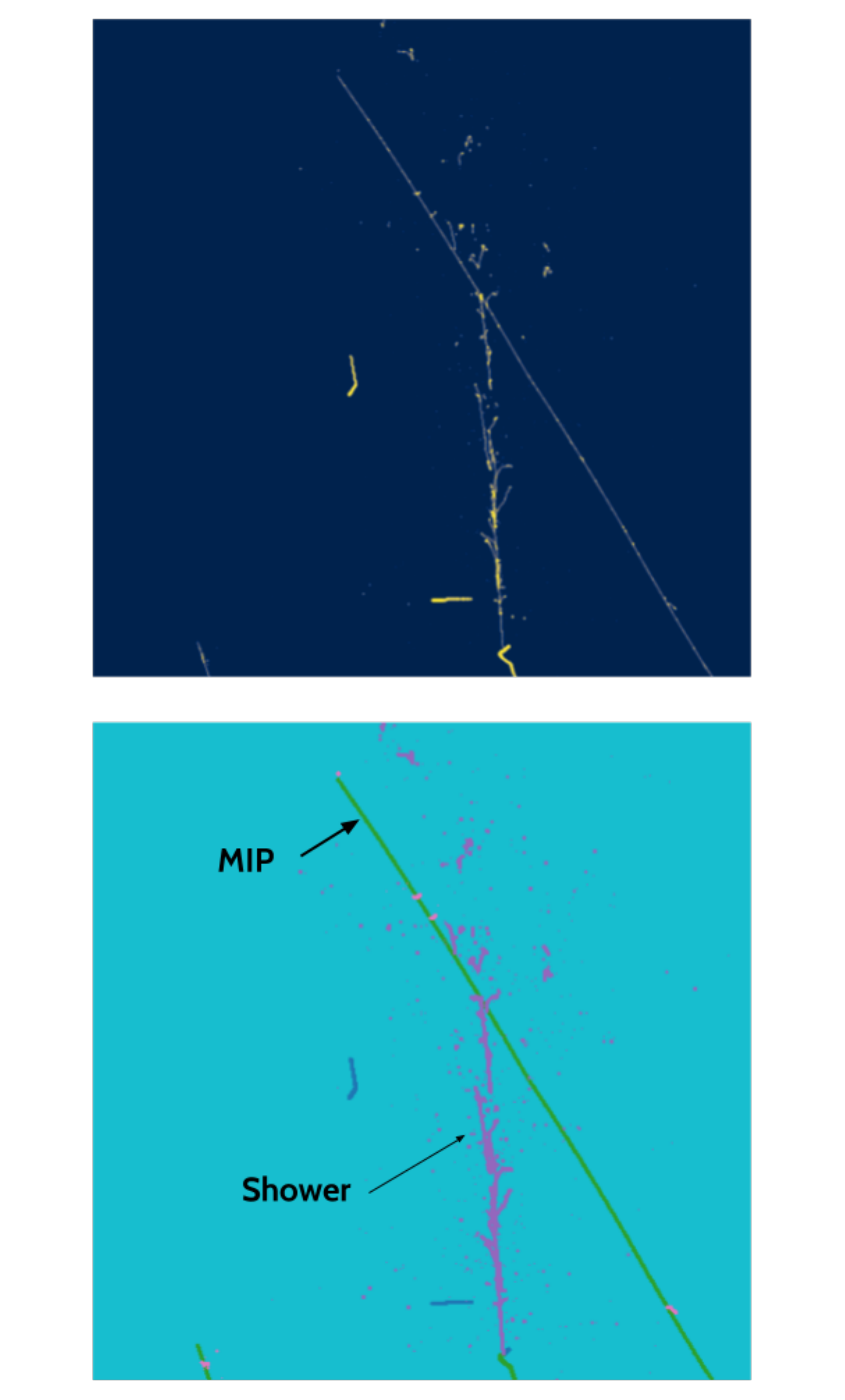}
    \caption{Top: energy depositions in the image, the pixel color corresponds to an energy scale. Bottom: labels, each color corresponds to a different class. Electromagnetic shower pixels are colored in purple and MIP pixels are in green.}
    \label{fig:viz_data}
\end{figure}
\begin{figure*}[t]
    \centering
    \includegraphics[width=0.98\textwidth]{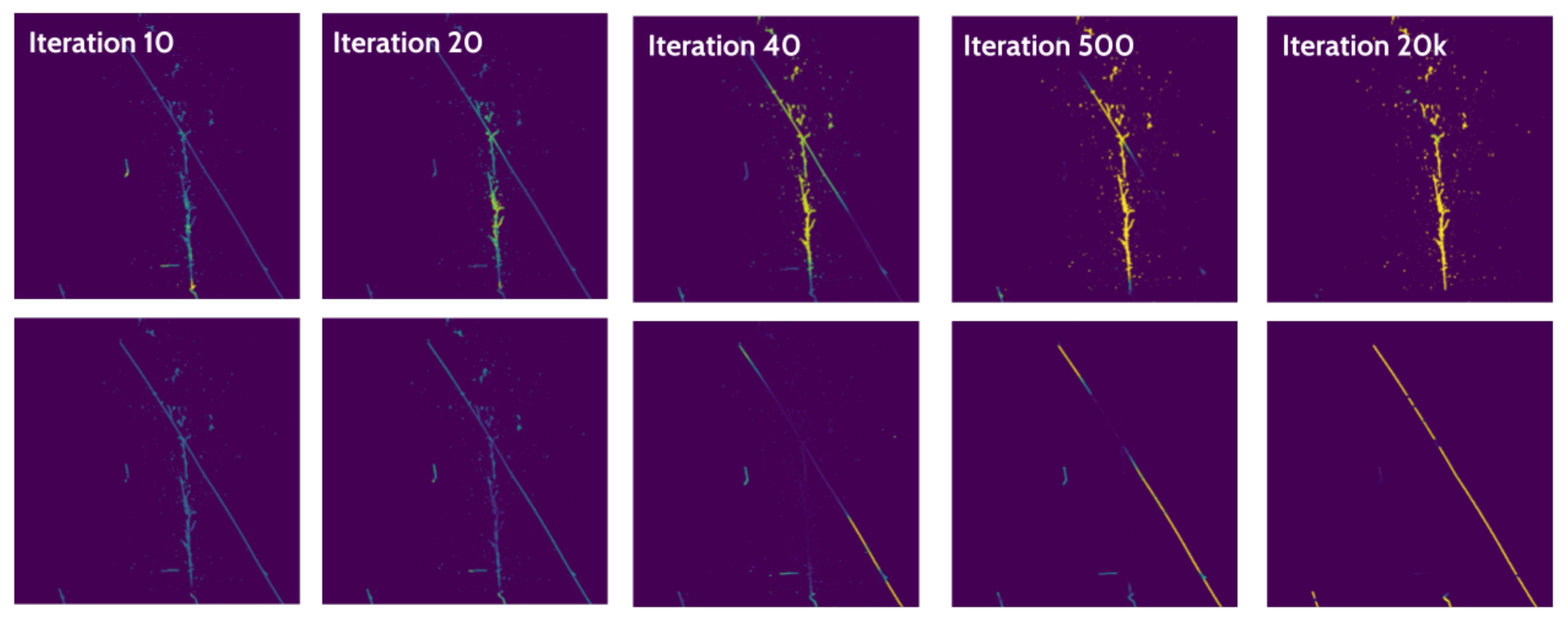}
    \caption{Dense U-ResNet evolution of softmax value for EM shower (top) and MIP (bottom) across training iterations.}
    \label{fig:viz_dense}
\end{figure*}
\begin{figure*}[t]
    \centering
    \includegraphics[width=0.98\textwidth]{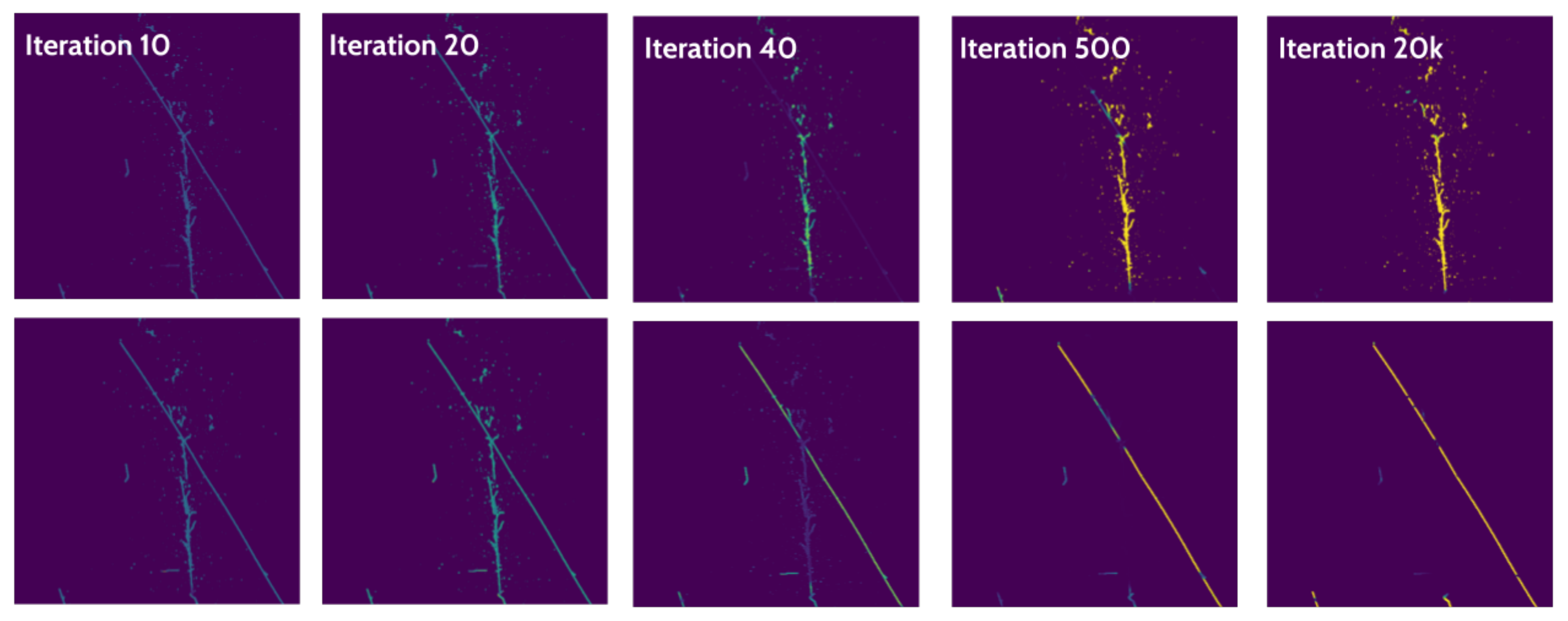}
    \caption{Sparse U-ResNet evolution of softmax value for EM shower (top) and MIP (bottom) across training iterations.} 
    \label{fig:viz_sparse}
\end{figure*}

Finally looking at the evolution of the softmax scores for different classes across iterations indicates that the sparse U-ResNet may be learning more uniformly over pixels than its dense equivalent. Figure~\ref{fig:std} shows how the standard deviation of the mean softmax value in the image evolves with the training iterations. The dense network results in a much higher variance. Their variances end up converging after about 1000 iterations. This observation is illustrated in Figure~\ref{fig:viz_data}, in which a MIP trajectory crosses an EM shower.  Figures~\ref{fig:viz_dense} and~\ref{fig:viz_sparse} compare how the softmax scores for track and shower particles change over training iterations between sparse and dense U-ResNet.  The difference appears most strikingly at the iteration 40, where the dense network is extremely confident in some pixels (yellow ones) and still very unsure about others (in dark blue), while the sparse one is increasing its confidence level much more uniformly across all pixels.

\subsection{Sparse U-ResNet Performance Variation}

\begin{figure}[t]
    \centering
     \includegraphics[width=0.48\textwidth]{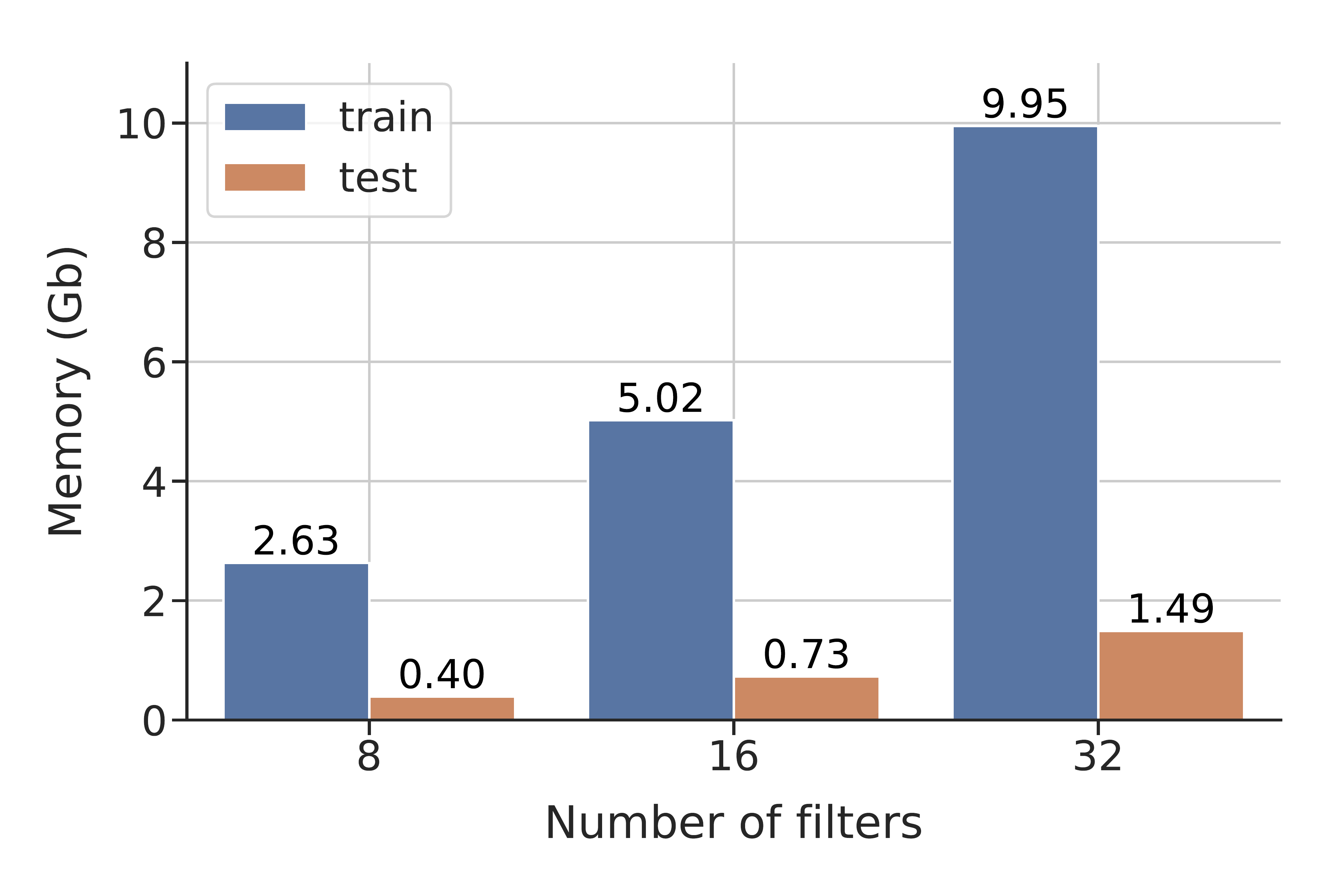}
     \caption{Memory usage of sparse U-ResNet with depth 6, 3D 512px images and a varying number of initial filters.}
     \label{fig:filters_usage_memory}
\end{figure}
\begin{figure}[t]
    \centering
    \includegraphics[width=0.48\textwidth]{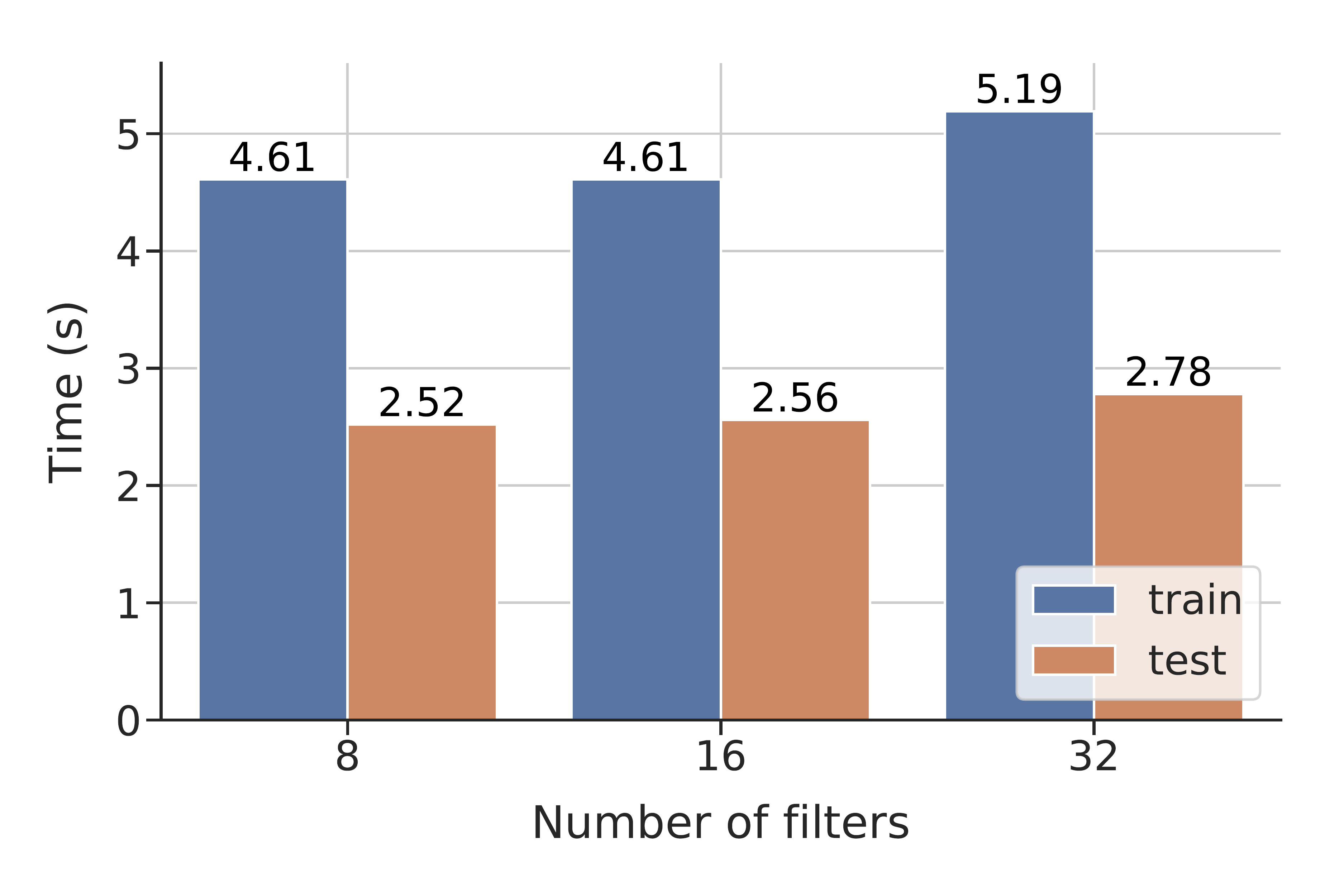}
    \caption{Computation wall-time of sparse U-ResNet with depth 6, 3D 512px images and a varying number of initial filters.}
    \label{fig:filters_usage_time}
\end{figure}

We study the influence of the two main parameters of the network architecture on performance and resource usage: depth (number of layers) and the number of filters in the first layer. Table~\ref{tab:sparse3d_performance}, Figure~\ref{fig:filters_usage_memory} and~\ref{fig:filters_usage_time} show the results of non-zero accuracy and computational resource usage respectively. The filter counts have a larger effect on achieving a higher accuracy while it also causes a linear increase in memory usage. The increase in the computation wall-time is only $\approx10$\% between 8 and 32 filter counts. 

\begin{table}[t]
    \centering
    \begin{tabular}{|l|c|c|c|}
        \hline
        Filters & 8 & 16 & 32 \\
        \hline
        Depth 6 & 98.94\% & 99.16\% & 99.23\% \\
        \hline
        Depth 5 & 98.86\% & 99.07\% & 99.06\% \\
        \hline
        Depth 4 & 98.74\% & 99.00\% & 99.07\% \\
        \hline
    \end{tabular}
    \caption{Comparison of the non-zero accuracy at inference time on the test set of 3D 512px images for sparse U-ResNet, for different depths and initial number of filters.}
    \label{tab:sparse3d_performance}
\end{table}

\begin{table}[t]
    \centering
    \begin{tabular}{|l|c|c|c|c|c|c|}
        \hline
         Test image & \multicolumn{3}{c|}{192px}  & \multicolumn{3}{c|}{768px}  \\
         \hline
         Train image & 192px & 512px & 768px & 192px & 512px & 768px \\
         \hline
         HIP & 96.0\% & 95.6\% & 93.7\% & 98.8\% & 99.0\% & 98.9\% \\
         \hline
         MIP & 96.2\% & 96.6\% & 95.4\% & 99.4\% & 99.7\% & 99.6\% \\
         \hline
         EM shower & 97.6\% & 96.9\% & 96.6\% & 99.5\% & 99.6\% & 99.7\% \\
         \hline
         Delta rays & 74.3\% & 76.7\% & 75.1\% & 85.9\% & 89.6\% & 90.1\% \\
         \hline
         Michel $e^{-}$ & 36.5\% & 42.6\% & 43.9\% & 62.6\% & 70.0\% & 70.4\% \\
         \hline
         \hline
         \textbf{Overall} & \textbf{98.0\%} & \textbf{98.1\%} & \textbf{97.7\%}  & \textbf{98.9\%} & \textbf{99.2\%} & \textbf{99.3\%} \\
         \hline
    \end{tabular}
    \caption{Class-wise non-zero accuracy. Comparing the performance of sparse U-ResNet with different 3D image sizes at train and test time. The batch size, the depth, and the initial number of filters are 64, 5, and 16 respectively.}
    \label{tab:scale_image}
\end{table}

Table~\ref{tab:scale_image} shows the result of comparing network class-wise non-zero accuracies for varying 3D image sizes at the train and test times.
For a given image size at train time, using a larger image size at test time systematically improves the performance.
This table also shows that the class-wise accuracy of delta rays and Michel electrons is lower than other classes across all image sizes.

\subsection{Mistakes Analysis}

One may consider that a poor performance may be partially due to particle trajectories being cut out at the recorded volume boundaries. We looked at the distribution of the fraction of mis-classified pixels as a function of their distance to the boundaries, which is defined as follows in $d$ dimensions: 
\begin{equation}
d(\{\mathbf{x_i}\}_{i=1,...,d}) = \min_{i=1...d} \min \mathbf{x_i}
\end{equation}
where $d$ runs from 1 to 3 for 3D data. In other terms, the distance of the pixel to the image boundaries is the distance from the pixel to the closest face of the cubic image boundaries. 

\begin{figure}[t]
    \centering
    \includegraphics[width=0.48\textwidth]{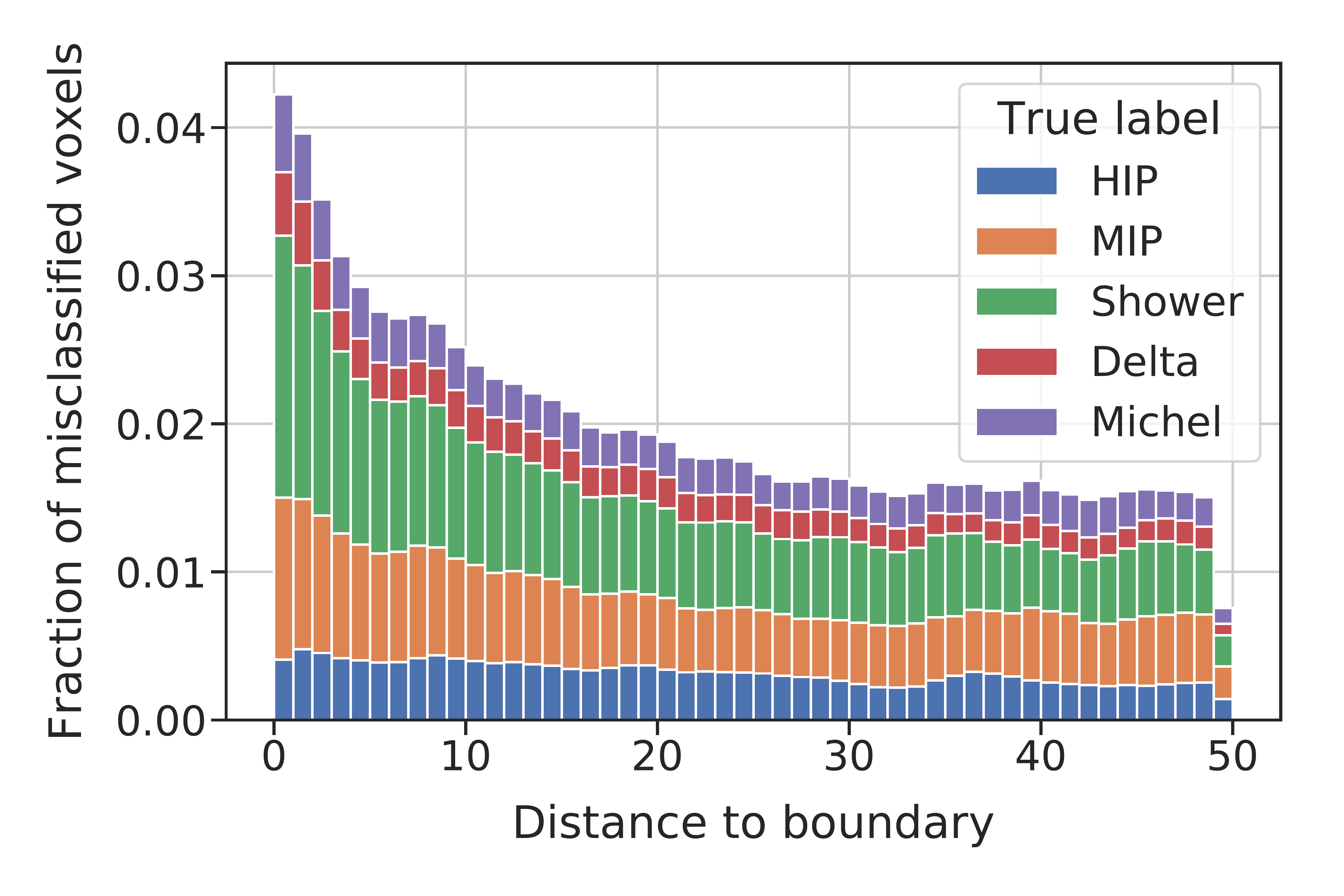}
    \caption{Fraction of misclassified pixels as a function of the pixel distance to the image boundaries [3D, 192px, 5-16].}
    \label{fig:misclassified}
\end{figure}

Figure~\ref{fig:misclassified} shows that, in general, pixels are more likely mis-classified near image boundaries as expected. We can see that, however, this is not clearly visible for Michel electrons and delta rays. Therefore this is not an explanation for the poor performance on these two classes. We investigated possible explanations beyond originally planned experiments, and report our findings in the following sections.

\begin{figure*}[t]
    \centering
    \includegraphics[width=0.27\textwidth]{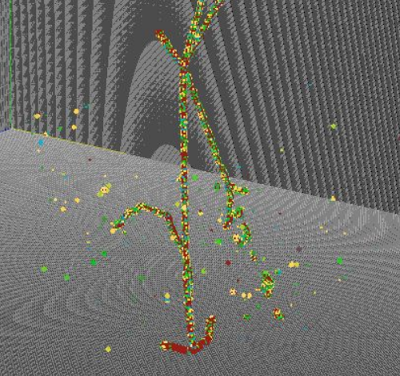}
    \includegraphics[width=0.27\textwidth]{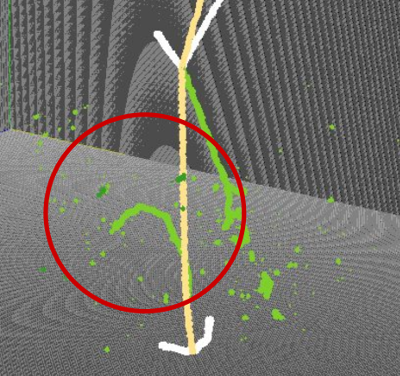}
    \includegraphics[width=0.27\textwidth]{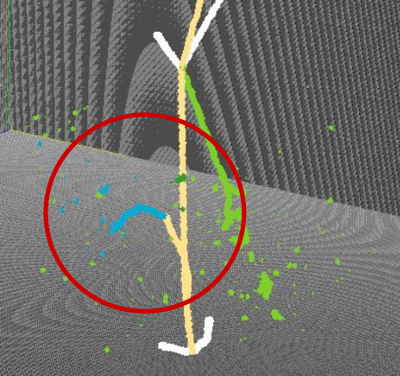}
    \\
    \includegraphics[width=0.27\textwidth]{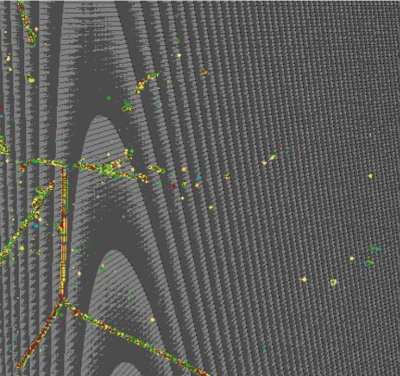}
    \includegraphics[width=0.27\textwidth]{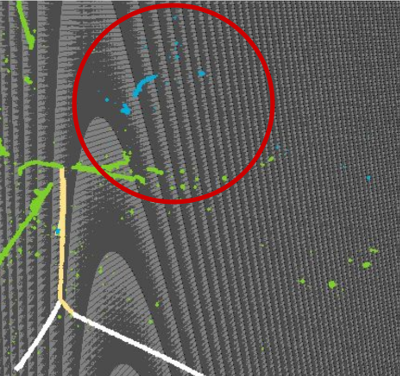}
    \includegraphics[width=0.27\textwidth]{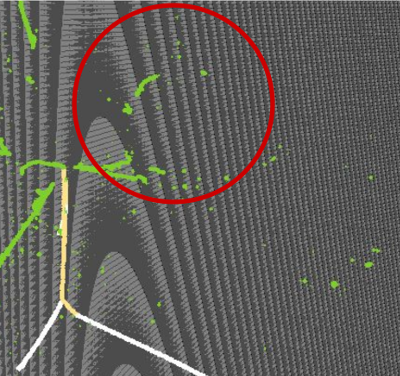}
    \\
    \includegraphics[width=0.27\textwidth]{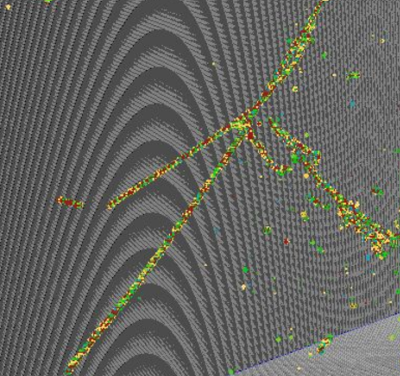}
    \includegraphics[width=0.27\textwidth]{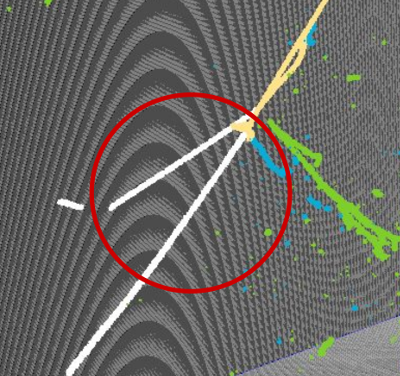}
    \includegraphics[width=0.27\textwidth]{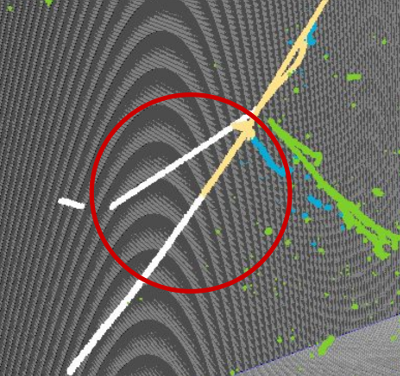}
    \\
    \includegraphics[width=0.27\textwidth]{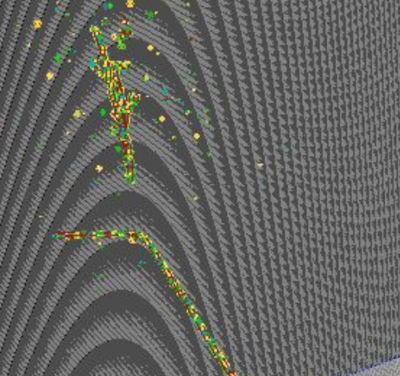}
    \includegraphics[width=0.27\textwidth]{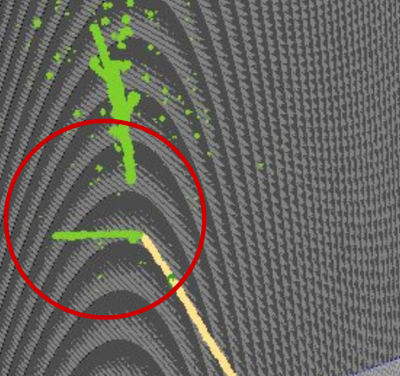}
    \includegraphics[width=0.27\textwidth]{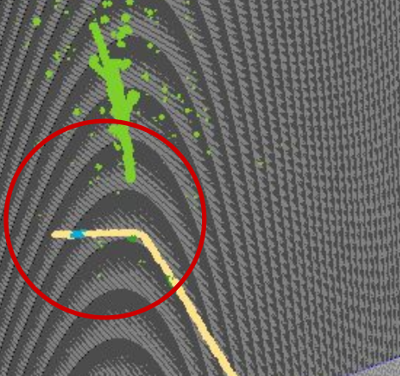}
    \caption{Typical mistakes of sparse U-ResNet [3D, 512px, 5-16]. Images are selected among the worst 0.05\% with respect to the non-zero accuracy metric. Mistakes are circled in red. Left column: data. Middle column: labels. Right column: predictions of the network. First row: an electromagnetic shower is mistaken for a Michel electron. Second row: a Michel electron is mistaken for an electromagnetic shower. Third row: a part of a HIP is mistaken for a MIP. Fourth row: a short shower is mistaken for a (MIP) track.}
    \label{fig:event_displays}
\end{figure*}

Figure~\ref{fig:event_displays} shows some 3D images of size 512px, which are examples of typical mistakes made by the sparse U-ResNet. This includes: Michel electrons mistaken for an electromagnetic shower and vice-versa, HIP mistaken for a MIP and the track-like beginning of a short EM shower mistaken for a MIP.

\section{Michel Electrons and Delta Rays}
\label{sec:discussion}

We propose two hypothetical explanations for low prediction accuracies on Michel and delta ray pixels by U-ResNet. The first is statistical imbalance in the fraction of pixels of each class: delta rays and Michel electrons represent each about 1~\% of the total pixels. The second is an ambiguous definition of these two classes: both Michel electrons and delta rays can emit gamma rays (e.g. Bremsstrahlung radiation) which appear to be indistinguishable from EM shower class. During training, we employed a softmax loss for classifying pixels under the assumption of exclusive class definitions, which may not hold for these classes. 

We implemented two changes in order to test our hypothesis. The first is a modification to the pixel labels used in our supervised training. For Michel electrons and delta rays, pixels are re-labeled as EM shower except for those that belong to a primary ionization trajectory, which carries distinctive features. Secondly we experimented a pixel-wise loss weighting factor to accommodate statistical imbalance across five classes. This allows U-ResNet to focus more on pixels with low statistics, inspired by attention mechanisms.

\begin{table}[t]
    \centering
    \begin{tabular}{|l|c|c|c|c|}
        \hline
         Train data & \multicolumn{2}{c|}{Regular} & Relabeled & Relabeled+Weights \\
         \hline
         Test data & Regular & \multicolumn{3}{c|}{Relabeled} \\
         \hline
         HIP & 98.0\% & 98.1\% & 98.1\% & 99.3\% \\
         \hline
         MIP & 99.4\% & 99.2\% & 99.4\% & 98.1\% \\
         \hline
         Shower & 99.4\% & 97.9\% & 99.2\% & 99.2\% \\
         \hline
         Delta rays & 85.7\% & 94.8\% & 96.0\% & 97.2\% \\
         \hline
         \textbf{Michel $e^{-}$} & \textbf{56.6\%} & \textbf{94.4\%} & \textbf{94.7\%}  & \textbf{95.7\%} \\
         \hline
         \hline
         Overall & 99.2\% & 99.2\% & 99.6\% & 99.1\% \\
         \hline
    \end{tabular}
    \caption{A comparison of class-wise nonzero accuracies between 3 flavors of sparse U-ResNet: regular, trained with relabeled dataset, and trained with both the relabeled dataset and the weighting scheme [3D, 512px, 5-32]. We also compare the performance of the regular sparse U-ResNet on a test relabeled dataset. }
    \label{tab:cw}
\end{table}

We train a sparse U-ResNet using the re-labeled dataset and optionally the pixel-wise loss weighting scheme. The results are presented in Table~\ref{tab:cw}. First, regardless of whether the U-ResNet was trained on the regular or relabeled dataset, the non-zero accuracy on Michel electrons increased by more than 40\%. This implies that the algorithm did learn the distinctive features of Michel electrons and delta rays without relabeling. Secondly, we see a slight improvement for delta rays and EM shower pixels by training with the re-labeled dataset. Finally, pixel-wise loss weighting further improved the accuracy of both Michel and delta ray classes as expected. 

\section{Michel Electron Reconstruction}
\label{sec:michel}
Finally we present a study on reconstructing Michel electron energy spectrum using the public simulation sample. Michel electron is one of well understood physics signals, and thus useful for detector calibrations. This analysis has been done by LArTPC experiments with real data including MicroBooNE and ICARUS~\cite{UBMichel,ICARUSMichel}. Our contribution is to show the first ML-based approach with quantification of both efficiency and purity of reconstructed signal.

\subsection{Reconstruction Method}
Our goal is to quantify the efficiency and purity of clustering Michel electron energy depositions by only using the primary ionization component of its trajectory. We use the 3D 512px images from the re-labeled sample presented in the previous section. After running the U-ResNet for semantic segmentation we isolate pixels that belong to each of the five classes. We run a common density-based spatial clustering algorithm DBSCAN~\cite{DBSCAN} to identify different predicted Michel electron clusters and MIP clusters, with parameters $\epsilon=2.8$ and $min_{Pts}=$ 5 and 10 respectively. We then select the candidate Michel electron clusters that are attached to the edge of a predicted MIP cluster. Here ``attached'' is defined as less than 1px distance between the nearest pixels of a Michel electron and MIP clusters. The ``edginess'' of a given pixel is evaluated by masking surrounding pixels within the radius of 15px, and making sure that the DBSCAN algorithm only finds one cluster when run over the remaining MIP cluster pixels.

\subsection{Performance Metrics}
After identifying candidate Michel electron clusters, we match each of them to a true Michel cluster by maximizing the overlap pixel count between true and predicted Michel cluster. We can then define several performance metrics. Let us define notations: $N_i^{\text{pred}}$ is the total number of pixels in the predicted Michel electron cluster $i$, $N_i^{\text{true}}$ the total number of pixels in the matched true Michel electron cluster $i$, and $N_i$ the number of pixels which belong to the intersection of both candidate and matched Michel electron clusters. Then we define clustering efficiency and purity as $N_i / N_i^{\text{true}}$ and $N_i / N_i^{\text{pred}}$ respectively.
Similarly if $N^{\text{true}}$ is the total number of true Michel electron clusters in the sample, $N_{\text{pred}}$ is the total number of candidate Michel electron clusters, and $N_{\text{pred}}^{\text{true}}$ is the number of matched candidate Michel electron clusters over the whole sample, then we define ID efficiency and purity as $N_{\text{pred}}^{\text{true}} / N^{\text{true}}$ and $N_{\text{pred}}^{\text{true}} / N_{\text{pred}}$ respectively.

\subsection{Results}
\begin{figure}[t]
    \centering
        \includegraphics[width=0.48\textwidth]{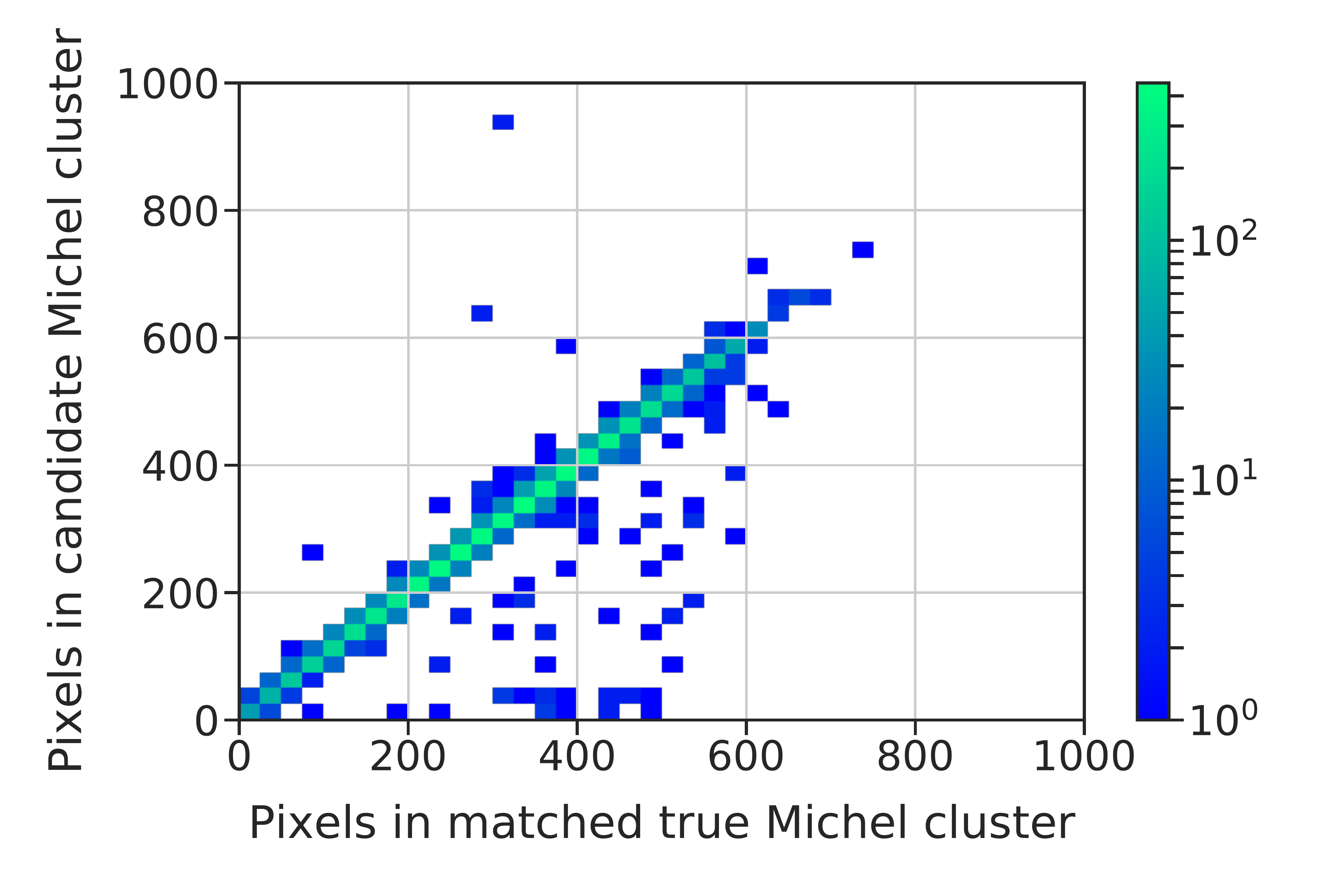}
    \caption{A comparison of pixel counts between the true and candidate Michel electron clusters. }
    \label{fig:num_true_pix_vs_num_pred_pix}
\end{figure}

Figure~\ref{fig:num_true_pix_vs_num_pred_pix} shows the pixel count of matched true Michel electron clusters against the pixel count of reconstructed Michel electron clusters. As expected, most of the clusters lie on the diagonal. The majority of strayed clusters are present below the diagonal and are under-clustered.

Table~\ref{tab:michel_eff} shows the evaluation metrics with and without an analysis quality cut, which requires reconstructed Michel electron clusters to contain 10 or more pixels. We find that 
89.8~\% of the reconstructed Michel electrons have both cluster efficiency and purity above 95\%.  MicroBooNE collaboration has published Michel electron reconstruction study with 2\% ID efficiency and 80-90\% ID purity where the focus of the analysis was to maximize the purity of the sample for accurate energy reconstruction~\cite{UBMichel}. The outcome of this study with the public simulation sample cannot be directly compared with others using real detector data because the public simulation sample lacks complicated detector effects. However, the results are compelling to show the promise of ML-based reconstruction approach.
\begin{table}[t]
    \centering
    \begin{tabular}{|c|c|c|}
        \hline
        Cut & None & 10 \\
        \hline
        Sample size & 6998 & 6961 \\
        \hline
        ID purity & 96.7~\% & 97.3~\% \\
        \hline
        ID efficiency & 93.9~\% & 93.4~\% \\
        \hline
        Cluster efficiency & 95.4~\% & 96.0~\% \\
        \hline
        Cluster purity & 95.5~\% & 96.0~\% \\
        \hline        
    \end{tabular}
    \caption{ID purity and efficiency as well as cluster purity and efficiencies of reconstructed Michel electrons. The sample size is the number of true positives. The cluster efficiency and purity are averaged over all reconstructed Michel electron clusters.}
    \label{tab:michel_eff}
\end{table}

\begin{figure}[ht]
    \centering
    \includegraphics[width=0.48\textwidth]{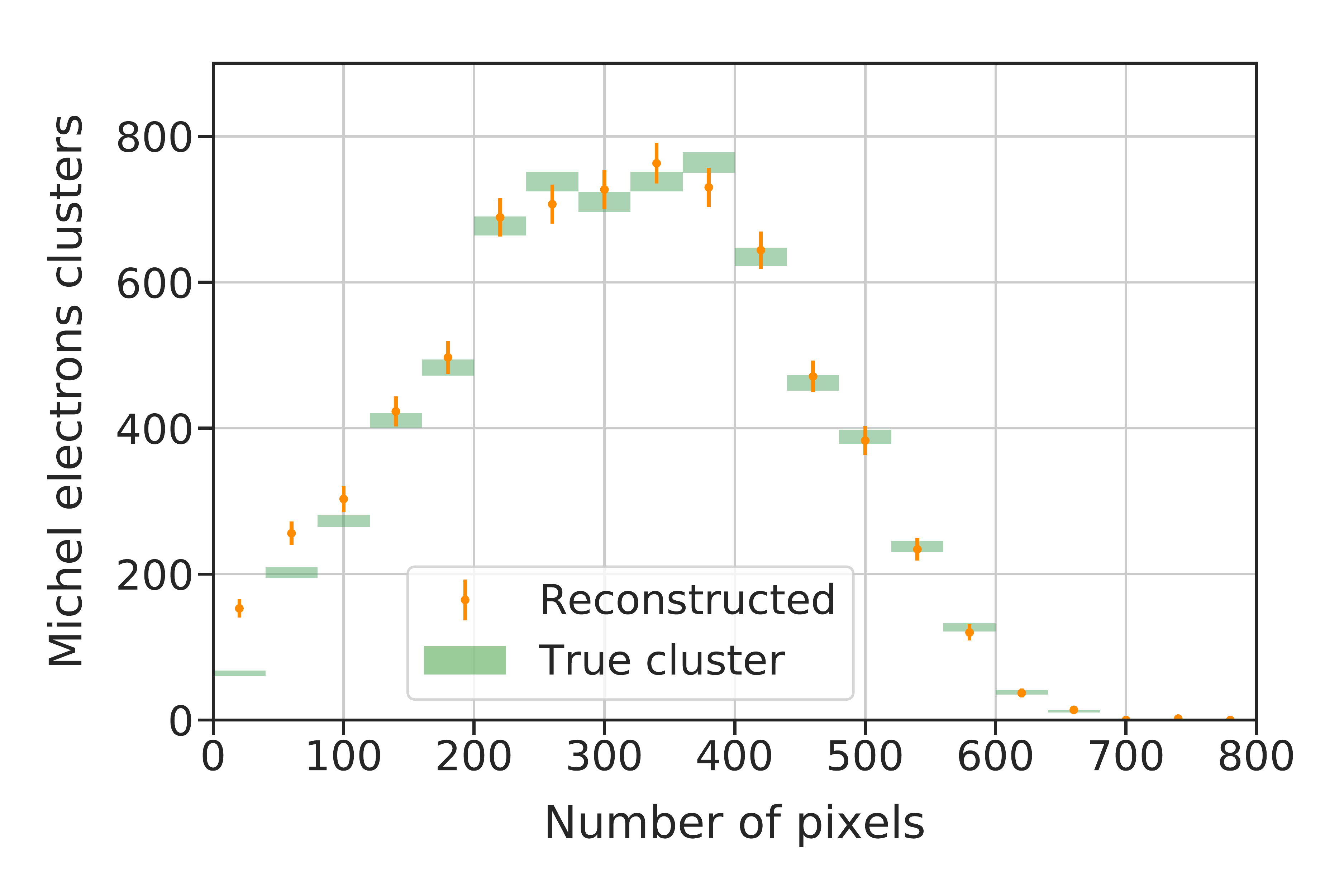}
    \caption{Energy spectrum of Michel electrons. The primary ionization energy is the energy of Michel after relabeling. The candidate Michel energy is the sum of pixel values predicted as Michel electron by the U-ResNet [3D, 512px, 5-32].}
    \label{fig:michel_spectrum}
\end{figure}

Finally, using the matched candidate Michel electrons, we compare the reconstructed and true energy distribution in the primary ionization component. Figure~\ref{fig:michel_spectrum} shows a reasonable agreement. In order to reconstruct the total true Michel electron energy, the reconstruction step needs to account for EM shower pixels resulting from Bremsstrahlung radiation as described in the MicroBooNE publication~\cite{UBMichel}. This is out of the scope of this paper.

\section{Conclusions}
\label{sec:conclusion}
In this paper, we demonstrated the strong performance of SSCN against our baseline dense CNN for LArTPC data reconstruction, specifically for the task of semantic segmentation to identify five particle classes at a pixel-level. We employed U-ResNet, an architecture pioneered by MicroBooNE collaboration, and showed that the implementation using SSCN makes a drastic improvement in the computational resource usage. For U-ResNet under the same condition of batch size 4 with 192px 3D images, SSCN reduces the computational cost in memory and wall-time at inference by a factor of 354 and 33 respectively as shown in Table~\ref{tab:dense_vs_sparse}. For 2D samples, using batch size 88, those reduction factors are 93 and 3.1 respectively. While a naive application of standard CNN for 3D data (e.g. the DUNE near detector) comes with prohibitive and extremely inefficient computational resource usage, we demonstrated that SSCN can mitigate such costs and generalize U-ResNet for 3D data samples without loss in the algorithm performance.

We presented the first demonstration of reconstructing Michel electron clusters, defined as the primary ionization component of a trajectory, using a primarily ML-based method. Our result using the public simulation sample shows a naive approach with DBSCAN on U-ResNet output can yield 93.9\% Michel electron identification efficiency with 96.7\% true positive rate. Pixel clustering efficiency for reconstructed Michel electrons is found to be 95.4\% with the purity of 95.5\%. In particular, 89.8\% of reconstructed Michel electrons are found to carry both the efficiency and purity of clusters above 95\%.

SSCN is a solution to address scalable CNN applications for LArTPC data, which is generically sparse but locally dense. Furthermore, SSCN is a generic alternative to dense CNN, and can be applied to tasks beyond semantic segmentation including image classification, object detection and more. We strongly recommend SSCN for any CNN applications that exist for LArTPC experiments including MicroBooNE, ICARUS, SBND, and DUNE.

\section{Acknowledgement}
This work is supported by the U.S. Department of Energy, Office of Science, Office of High Energy Physics, and Early Career Research Program under Contract DE-AC02-76SF00515.

\bibliography{main}
\end{document}